# Large/small eddy simulations: *a posteriori* analysis in high Reynolds number isotropic turbulence

Chang Hsin Chen,[a] Arnab Moitro,[b] and Alexei Y. Poludnenko[c]

*School of Mechanical, Aerospace, and Manufacturing Engineering, University of Connecticut, Storrs, CT 06028, USA*

(*Electronic mail: alexei.poludnenko@uconn.edu)

(Dated: 8 December 2025)

While direct numerical simulations (DNS) are the most accurate method for studying turbulence, their large computational cost restricts their use to idealized configurations and to Reynolds numbers well below those found in practical systems. A recently proposed method, Large/Small Eddy Simulation (L/SES), aims to overcome this limitation while still providing the solution fidelity comparable to that of DNS. L/SES represents a pair of coupled calculations: a lower-fidelity Large Eddy Simulation (LES), which captures the large-scale flow structure, and a high-fidelity Small-Eddy Simulation (SES) targeting a sub-region of interest of the LES, in which the small-scale dynamics is fully resolved. In this study, we demonstrate the accuracy and performance of L/SES in large Reynolds-number homogeneous isotropic turbulence (HIT) up to Taylor-scale Reynolds number $\mathrm{Re}_\lambda \approx 600$. Turbulence properties obtained with L/SES are shown to be in close agreement with the literature, both in terms of global characteristics, such as kinetic energy spectra and dissipative anomaly, as well as small-scale properties, such as higher-order moments of the velocity gradients up to the $10^{\mathrm{th}}$ order and probability density functions of the intermittent quantities. Also using simulations of HIT, we systematically investigate various method parameters and determine their optimal converged values. Finally, we discuss the computational cost of L/SES and demonstrate that it is $\approx 3$ orders of magnitude lower than for a traditional DNS at the highest Reynolds number considered here. This highlights the potential of L/SES as a discovery tool, which brings high-fidelity simulations of realistic flows into the realm of feasibility.

## I. INTRODUCTION

Turbulence is an inherently multi-scale flow phenomenon. Its complex nonlinear dynamics is fully manifest when the largest energy containing scales become sufficiently greater than the smallest scales, on which turbulent kinetic energy dissipates. The Reynolds number - the figure of merit reflecting such complexity of turbulent flows - can range from a few thousand in small-scale engineering systems on Earth to $\sim 10^{13} - 10^{16}$ in stellar interiors during core-collapse or Type Ia supernovae explosions[1-3]. Direct numerical simulations (DNS), which resolve all fluid scales of a turbulent flow, are not feasible for any practically relevant Reynolds numbers even using modern supercomputing resources. The reason for this is that the computational cost of a DNS per unit time grows rapidly with the Taylor-scale Reynolds number[4] as $\sim \mathrm{Re}_\lambda^{9/2}$. The largest $\mathrm{Re}_\lambda$ achieved thus far[5-8] in the DNS of homogeneous isotropic turbulence (HIT) are $\sim 1300 - 2500$ requiring grid sizes $N^3 = 12,288^3$ to $32,768^3$. Such extreme calculations are possible only on the state-of-the-art exascale computational platforms, and yet Reynolds numbers that they can reach are still orders of magnitude below those representative of most engineering and natural systems. Furthermore, their extreme cost limits the exploration of the physics of turbulence beyond the idealized configurations, such as homogeneous isotropic turbulence (HIT).

In a recent paper, Moitro, Dammati, and Poludnenko[9] (hereafter, MDP; also see Chen, Moitro, and Poludnenko[10]) proposed a new method, Large/Small Eddy Simulation (L/SES), to enable simulations of turbulent flows with the DNS-level fidelity but at a fraction of the cost of a DNS. The primary goal of this approach is to allow first-principles exploration of turbulence at Reynolds numbers beyond those, which can be achieved with modern DNS.

The L/SES method is based on two coupled simulations. The first one is a lower-fidelity calculation, which in the current formulation is a Large Eddy Simulation (LES). It is intended to capture with sufficient accuracy the large-scale dynamics of the entire flow of interest. Such LES is coupled with a high-fidelity Small-Eddy Simulation (SES), which fully resolves the small scales in a sub-region of interest of the LES. Presently, such coupling is one-way. It is achieved by filtering and interpolation of the LES flow field at a suitable filter scale. The resulting forcing data effectively provide a large-scale model, which is used to nudge the SES solution to recover the effect of the energy-containing scales of the flow. As such, L/SES replaces a single calculation, i.e., DNS, which must capture the full range of scales, with two calculations, which target two separate sub-ranges individually: one from the largest scales down to the filter scale, which is typically comparable to the Taylor micro-scale, and another from the filter scale down to the dissipative scale.

The L/SES method was described in detail in MDP. At the same time, that study had three key limitations, which we seek to address in the current paper. First, accuracy of the method was analyzed only in the *a priori* sense using SES calculations, for which the large-scale forcing fields were obtained by filtering the fully resolved DNS, rather than LES. Such tests were aimed to determine the theoretical limit of accuracy of the method in a situation when all scales, both large and small, are accurate. Furthermore, this allowed the analysis of the SES solution accuracy not only in a statistical sense, but through a direct point-by-point comparison with the DNS flow field, which the SES was intended to recover. Here, we aim to perform the *a posteriori* analysis of the accuracy of the L/SES in a practically relevant setting when the large-scale dynamics is obtained from the LES. Furthermore, to consider the worst-case scenario, in all calculations described below, we will rely on implicit LES[11], which do not employ any subgrid-scale models, and thus which have the lowest accuracy of the small-scale flow solution. The goal is to determine how sensitive the SES is to the errors in the LES flow field.

---
[a] Currently at Department of Mechanical Engineering, Texas Tech University, Lubbock, TX 79409, USA

[b] Currently at School of Engineering, Newcastle University, Newcastle upon Tyne NE1 7RU, UK

[c] Also at Department of Aerospace Engineering, Texas A&M University, College Station, TX 77843, USA

Second, the L/SES method relies crucially on a number of free parameters, such as the filter scale, LES grid resolution, size of the SES domain, frequency of the LES flow-field sampling for SES forcing, etc. In MDP, the rationale for the choice of these parameters was discussed either based on the qualitative theoretical arguments or idealized, one-dimensional (1D) tests. Here we systematically explore the effect of each parameter through several series of three-dimensional (3D) calculations of HIT, in which each parameter is varied individually. The optimal choice of each parameter is determined by considering the solution convergence in terms of a range of metrics, from the normalized dissipation to the higher-order moments of the velocity-gradient statistics.

Third, in MDP, the L/SES method was demonstrated for HIT flows only at relatively low Reynolds numbers $\mathrm{Re}_\lambda \sim 100$, which are also easily accessible using classical DNS. Since the properties of a turbulent flow, for instance its intermittency, vary with the Reynolds number, it is important to demonstrate the ability of the L/SES to recover the small-scale turbulence structure correctly at more representative $\mathrm{Re}_\lambda$. It is also equally important to show the computational efficiency of the method, and in particular how it varies with $\mathrm{Re}_\lambda$. To this end, we perform a series of L/SES calculations for $\mathrm{Re}_\lambda$ in the range $\sim 100 - 600$. Since here we consider a well-studied canonical flow configuration, namely HIT, a wide variety of turbulence characteristics obtained for these Reynolds numbers can be directly compared with the previously published theoretical and computational results in the literature to assess the accuracy of the L/SES method.

Turbulence characteristics that we consider in this study, on one hand, represent the global properties of the flow, including the integral scale and Taylor micro-scale, total specific turbulent kinetic energy (TKE) and its spectral density, and normalized ensemble-average dissipation. On the other hand, complex nonlinear dynamics of turbulence is predominantly associated with the velocity derivatives, rather than velocities themselves. This is reflected in the probability density functions (PDF) of various velocity-derivative-based quantities, such as enstrophy or dissipation, as well as various statistical moments of the velocity derivatives. Therefore, in the analysis below, particular emphasis is placed on the assessment of the ability of the L/SES to recover accurately the details of the velocity-derivative statistics, and its dependence on the Reynolds number.

Previous studies have reported that the PDF of the velocity-derivative statistics are nearly Gaussian at low $\mathrm{Re}_\lambda$.[12,13] With increasing Reynolds numbers, however, the PDF depart from the Gaussian distribution and exhibit strong intermittency.[12,14,15] Both the exponential[16] and power-law[5] functions have been used to describe the long tails of the enstrophy and dissipation PDF for different $\mathrm{Re}_\lambda$. Regardless of different analytical fits, studies show that the PDF of enstrophy and dissipation demonstrate a clear dependence on $\mathrm{Re}_\lambda$.[17]

The $\mathrm{Re}_\lambda$ dependence was also found for the moments of the velocity-derivative statistics. For instance, the power-law dependence on $\mathrm{Re}_\lambda$ was shown for the flatness of the longitudinal velocity derivatives[14,18] More recently, theoretical description was proposed for the higher-order moments of the velocity-derivatives.[19,20] In particular, it was suggested that such higher-order moments remain close to a constant at low $\mathrm{Re}_\lambda$ and transition to a $\mathrm{Re}_\lambda$ scaling at high Reynolds numbers when the flow becomes non-Gaussian. The proposed theoretical relations were found to be consistent with the DNS data.[19,21]

Such $\mathrm{Re}_\lambda$ dependence of the velocity-gradient statistics, predicted by the theory and observed in the DNS, provides a very sensitive measure of the complex small-scale dynamics of turbulence. Therefore, the ability to capture such statistics in detail allows us to assess very precisely the accuracy of the L/SES solution, and more specifically of the small-scale turbulence dynamics targeted by the SES.

With these goals in mind, the paper is structured as follows. We start with the overall description of the numerical method used in the calculations, along with a summary of the LES and SES simulation setup (§ II). Next, we describe the analysis of the L/SES parameters as well as the numerical studies used to establish the optimal choice for their values (§ III). Based on these optimal values, next we present a suite of L/SES calculations for a range of $\mathrm{Re}_\lambda \sim 100 - 600$, along with the detailed analysis of the accuracy of various flow characteristics described above (§ IV). Computational cost of the L/SES and its $\mathrm{Re}_\lambda$ dependence are described in § V. Finally, we conclude with the discussion of the key findings (§ VI).

## II. LARGE/SMALL EDDY SIMULATION METHOD

Detailed description of the L/SES method can be found in MDP. Here we summarize the key steps of the algorithm, along with a few modifications[10] made in the current study compared to the original method described in MDP.

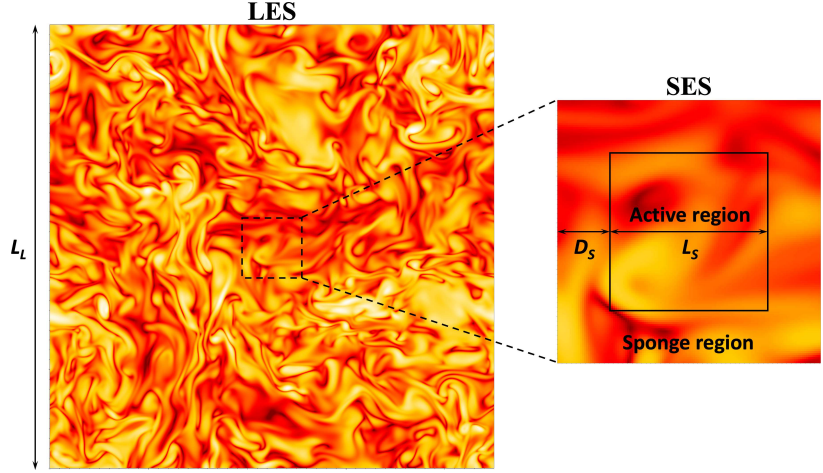

FIG. 1. Schematic of the L/SES computations. LES is performed in a triply-periodic domain. SES domain represents the central, cubic sub-region of the LES domain. See text for further description of the active and sponge regions of the SES.

## A. Governing equations and numerical method

Both LES and SES solve compressible Navier-Stokes equations

$$\frac{\partial \rho}{\partial t} + \frac{\partial \rho u_j}{\partial x_j} = 0, \tag{1}$$

$$\frac{\partial \rho u_i}{\partial t} + \frac{\partial \rho u_i u_j}{\partial x_j} = -\frac{\partial p}{\partial x_i} + \frac{\partial \sigma_{ij}}{\partial x_j} + f_i, \tag{2}$$

$$\frac{\partial \rho E}{\partial t} + \frac{\partial (\rho E + p) u_j}{\partial x_j} = \frac{\partial}{\partial x_j}\left(\kappa \frac{\partial T}{\partial x_j}\right) + \frac{\partial u_i \sigma_{ij}}{\partial x_j} + u_j f_j, \tag{3}$$

where $\rho$ is the density, $u_i$ is the velocity along the $i^{\text{th}}$ direction, $p$ is the pressure, $E = e + u_j u_j / 2$ and $e$ are the total and internal energy per unit mass, respectively, and $T$ is the temperature. Thermal conductivity $\kappa = 6.234 \times 10^3$ erg/(s cm K). The viscous stress tensor is

$$\sigma_{ij} = 2\mu S_{ij} - \frac{2}{3}\mu S_{kk}\delta_{ij}, \tag{4}$$

where $S_{ij} = \frac{1}{2}(\frac{\partial u_i}{\partial x_j} + \frac{\partial u_j}{\partial x_i})$, $\delta_{ij}$ is the Kronecker delta, $\mu = \rho\nu$ is the dynamic viscosity coefficient, and $\nu$ is the kinematic viscosity coefficient, the value of which is specified for each calculation below. Note, in eq. (4), we assume equal bulk and shear viscosity coefficients. Throughout this study, calorically perfect ideal gas is used with the equation of state $p = (\gamma - 1)\rho e$, where $\gamma = 1.197$ is the specific heat ratio. Finally, $f_i$ represents volumetric forcing, which has different meaning in the LES and SES calculations as described below.

Flow equations (1)-(3) are solved on a uniform, Cartesian grid using the code `Athena-RFX` - a fully compressible, massively parallel, numerical solver[9,22,23]. The code implements a higher-order, fully conservative, Godunov-type integration method with the unsplit corner transport upwind (CTU) algorithm[24,25]. The integration scheme uses piecewise parabolic method (PPM) for spatial reconstruction[26], along with an approximate nonlinear Harten–Lax–van Leer contact (HLLC) Riemann solver. The overall scheme is 3$^{\text{rd}}$-order accurate in space and 2$^{\text{nd}}$-order accurate in time. Further details of the integration algorithm and its implementation can be found in Gardiner and Stone[27] and Stone et al.[23], while further details of the a priori tests of the L/SES method with this code are provided in MDP.

## B. Large eddy simulations

In this study, the LES calculations do not include any explicit subgrid-scale (SGS) model, and thus they represent implicit LES.[11] We specifically chose this type of LES to analyze the resulting accuracy of the coupled L/SES approach in the case when the LES solution has the lowest accuracy of the small scales, which are dominated by the numerical dissipation. The main premise of the L/SES is that the low-fidelity calculation must provide only the reliable large-scale flow structure, and implicit LES allows us to test the validity of this premise.

LES calculations represent statistically stationary, homogeneous, isotropic turbulence (HIT) in a triply periodic domain (Fig. 1). In order to achieve such quasi-steady flow, classical spectral turbulence forcing, represented by the quantity $f_i$ in eqs. (1)-(3), is used to inject kinetic energy into the flow at the scale of the LES domain size $L_L$ at a constant rate $\epsilon_{inj}$.[28,29] The values of both $L_L$ and $\epsilon_{inj}$ can be set arbitrarily, and in all calculations described below $L_L = 0.45$ cm and $\epsilon_{inj} = 5.29 \times 10^9$ erg/g·s. Hereafter, the subscripts $L$ and $S$ denote the LES and SES quantities, respectively.

The key parameter of the calculations described below is the desired nominal Taylor-scale Reynolds number

$$Re_{\lambda^*} = \frac{u^*_{rms}\lambda^*}{\nu}. \tag{5}$$

Hereafter, asterisk indicates the predicted reference values of the corresponding quantities, which were calculated using the input parameters of the simulations. The predicted value of the Taylor microscale

$$\lambda^* = \left(15 \frac{\nu}{\epsilon_{inj}}\right)^{1/2} u_{rms}^*. \tag{6}$$

Since values of the outer turbulence scale and energy injection rate are set, the equilibrium value of $u_{rms}^*$, which is necessary to calculate $\mathrm{Re}_\lambda^*$ and $\lambda^*$, can be estimated using a theoretical expression for the normalized dissipation rate[30] (also see eq. 24 below). We find that $u_{rms}^*$ obtained through this procedure is within $\lesssim 10\%$ of the actual time-averaged value of $u_{rms}$ in a statistically stationary turbulence.

The forcing mechanism is described in detail in Poludnenko and Oran[22] and MDP and is only briefly described here. First, velocity perturbations for each velocity component are initialized in the Fourier space at the lowest wavenumber as independent realizations of a Gaussian random field. The non-solenoidal component is then removed to ensure that the perturbation field remains divergence-free. These perturbations are then transformed to physical space and normalized to ensure a constant prescribed rate of energy injection into the domain $\epsilon_{inj}$. The net momentum is also subtracted from the resulting velocity perturbation field $\delta u_i$ before it is directly added to the velocity field $u_i$ at each time step $n$, i.e., $u_i^{n+1} = u_i^n + \delta u_i^n$. The overall velocity perturbation pattern is regenerated at periodic time intervals to promote flow isotropy and homogeneity. Such forcing is similar to other approaches typically used in the DNS studies of the steady HIT turbulence[28,31]. It allows one to obtain a quasi-stationary, large-scale turbulent flow field with well characterized properties, which can be directly compared to other DNS of HIT in the literature.

LES calculations are initialized with a synthetic turbulent velocity field, which has only a solenoidal component and an ideal $\propto k^{-5/3}$ spectral kinetic energy distribution. The initial velocities are normalized to ensure that the total kinetic energy in the domain at $t = 0$ is approximately equal to its equilibrium value in the fully developed turbulent flow $(u_{rms}^*)^2/2$. After that, turbulence in the LES domain is allowed to reach a statistically stationary state over $\approx 7$ integral-scale eddy turnover times defined as

$$\tau_{ed} = \frac{l}{u_{rms}}. \tag{7}$$

Only the large-scale flow component from the LES is injected into the SES. Therefore, the small scales below a certain scale $\Delta$ must be removed by an explicit low-pass filter, since such scales may be affected by the lack of resolution as well as any inaccuracies in the SGS models (if such models are used in the LES). As described in MDP, it is critical for the filter employed to satisfy several important properties. In particular: (i) it must be spectrally sharp to ensure that only small scales $< \Delta$ are removed and there is minimal effect on the larger scales; (ii) it must be non-dispersive and it must introduce zero phase shift into the filtered data, which is important for the stability of the overall method; (iii) it must be implemented in physical space and it must be spatially local, which is important when applied to non-periodic domains and complex geometries; and finally (iv) it is desirable for the filter to be commutative with the derivative operator if such a filter were to be also used as part of the explicitly filtered LES solution.

In the current implementation, we use the 8th-order version of the low-pass differential filter[9,32] given by

$$\phi = \overline{\phi} + (-1)^{n/2} a^n \nabla^n \overline{\phi}, \tag{8}$$

where $\phi$ and $\overline{\phi}$ are the input and filtered quantities, respectively, $a = \Delta/\sqrt{40}$, and $n = 8$ is the order of the filter. The criteria for choosing the filter scale $\Delta$ will be discussed below. The density, $\rho$, and momentum, $\rho u_i$ (both conserved quantities), as well as pressure, $p$, are filtered. In this work, the filter is applied to the entire LES domain, and not just to the sub-region selected for the SES (cf. Fig. 1). Since the LES domain is periodic, this eliminates any boundary condition difficulties in applying the filter. In more realistic flows, which are not periodic, filter may be applied only to the subregion of interest. This would also reduce the overall computational cost of the method.

Since both the spatial and temporal resolution of the filtered LES data are much lower than those of the SES, LES data must be spatially and temporally interpolated before being injected into the SES. It is critical for spatial interpolation not to introduce any small-scale noise. Unlike the MDP, where we used spectral interpolation, in this work, all filtered LES quantities are spatially interpolated to the SES resolution directly in physical space using trigonometric interpolation. In 1D, such interpolation is expressed as[33]

$$y(x_j) = \sum_{i=0}^{N_0-1} y_{0,i}(x_{0,i}) \sin\left(N_0 \frac{x_j - x_{0,i}}{2}\right) \cot\left(\frac{x_j - x_{0,i}}{2}\right), \tag{9}$$

where $x_{0,i}$ is the coordinate of the $i$th point on the coarse grid with value $y_{0,i}(x_{0,i})$, $N_0$ is the total number of points on the coarse grid, and we seek the interpolated value $y(x_j)$ at the coordinate $x_j$ of the $j$th query point on the fine grid. This procedure is successively applied along all three dimensions to obtain the 3D interpolated data. Similar to spectral interpolation, this method ensures that the interpolation process does not create any artificial high-wavenumber noise. At the same time, it is implemented in physical space, and thus it does not require a Fourier transform of the filtered LES data. In addition, unlike spectral interpolation, it can be applied only in the sub-region of interest that will be captured in the SES, which reduces the computational cost of the interpolation.

Finally, since here we adopt an 'offline' approach, LES data are stored at discrete time instants, and its filtering and spatial interpolation are performed as a post-processing step. The choice of the time interval $\Delta t_L$ between the individual LES snapshots is discussed in further detail below.

## C. Small eddy simulations

To perform a fully-resolved coupled SES, a sub-region of the LES domain is selected. In the case of HIT considered in this work, any arbitrary region in the homogeneous flow can be chosen without loss of generality. In the calculations discussed here, the central cubic region of size $L_S$ is chosen for the SES domain, similar to MDP.

Initial state in an SES is set using the filtered and interpolated LES data. In principle, an arbitrary LES instant can be chosen to initialize the SES, though in all calculations discussed below, SES is started at $t = 6.9\tau_{ed}$ in the LES. Equations (1)-(3) are then evolved with the forcing based on the LES data. Unlike MDP, which modified the velocity field, here forcing is applied to the momentum at every time step $n$

$$(\rho u_i)_S^{n+1} = (\rho u_i)_S^n + \overline{(\rho u_i)}_L^n - \overline{(\rho u_i)}_S^n. \qquad (10)$$

Here, $\overline{\rho u_i}$ represents filtering using the same filter and the same filter scale $\Delta$ both in the LES and SES. Following MDP, $\overline{(\rho u_i)}_L^n$ is the filtered and spatially interpolated LES data, which is also linearly interpolated in time to the SES step $n$.

Forcing given by eq. (10) provides nudging of the large SES scales $> \Delta$ to the 'accurate' large-scale flow obtained from the LES, while at the same time minimally affecting the smaller SES scales $< \Delta$. It is volumetric in nature, and it is similar to the spectral nudging method used in weather modeling[34] (see MDP for further discussion of the similarities and differences between the L/SES method and analogous approaches used in weather modeling and data assimilation). Thus, the overall L/SES method splits the responsibility of recovering the large and small flow scales between the LES and SES calculations, respectively, with the LES and SES being coupled via eq. (10) thus resulting in a multi-fidelity approach.

Since SES focuses only on a small sub-region of the LES, its domain is not periodic and instead the SES boundary conditions are set based on the LES data. In order to ensure that the SES solution is fully compatible with the LES boundary values at all times, an additional sponge region is introduced outside the active SES domain (cf. Fig. 1; also see MDP for further details). In this region, the SES solution is relaxed to the LES boundary conditions using a 2nd-order polynomial function

$$\phi'(x) = \left(1 - \left(\frac{x}{D_s}\right)^2\right)\phi_S(x) + \left(\frac{x}{D_s}\right)^2 \overline{\phi}_L(x), \qquad (11)$$

where $x$ is the distance from the center of a given cell in the sponge region to the nearest outer cell of the active region, and $\phi_S(x)$ and $\overline{\phi}_L(x)$ are the values of a given conserved variable in the SES and filtered LES solutions, respectively, at the same position $x$. The $D_s$ is the width of the sponge region, discussed further below.

In MDP, quantitative comparison of the quadratic and exponential relaxation functions was performed using an idealized, synthetic, multi-modal 1D signal. It was found that both provide comparable accuracy of the relaxed solution. At the same time, here we chose to use the quadratic relaxation function given by eq. (11) since it ensures that the flow solution in the sponge region exactly matches the SES flow field at the inner boundaries of the sponge region and the filtered LES solution at the outer boundaries.

Thus, the SES domain shown in Fig. 1, in which eqs. (1)-(3) are solved, is comprised of the active and sponge regions, with the sponge region relaxation applied after each time step. At the end, only data from the inner active region can be used for analysis, while the data from the sponge region is discarded.

Finally, in MDP, an additional buffer region was introduced outside the sponge region to address the difficulty with the treatment of a boundary condition in the solution of the filtering eq. (8). This equation, which is effectively a higher-order nonhomogeneous Helmholtz equation, requires a boundary condition at infinity, which is not possible in a computation. A buffer region allows one to separate the boundary condition of the filtering equation from the boundary of the SES computational domain, thereby minimizing the error invariably introduced into the solution by the filtering operation. At the same time, such buffer region increases both the memory requirements and the overall computational cost associated with filtering in a region much larger than the SES domain. Therefore, in contrast with MDP, computations described here do not contain a buffer region, and instead a one-sided, 2nd-order finite difference operator was used to discretize eq. (8) near the outer boundaries of the sponge region. This allowed the filter scale to remain constant near the boundaries, albeit at the expense of a reduced filter order, and also increased the computational efficiency. The boundary values of the overall SES domain were again set using the interpolated LES data. We analyze the accuracy of this approach in detail below. Finally, we note that avoiding the buffer would have an added benefit of allowing the L/SES approach to be applied to wall-bounded flows, in which the physical domain boundary must be a part of the SES computational domain.

## III. KEY PARAMETERS OF THE L/SES METHOD

The primary parameters of an L/SES calculation are the LES domain size, $L_L$, and the SES grid resolution, $\Delta x_S$. These are directly related to the outer and inner scales of the turbulent flow of interest, and as such their values are dictated by the physical properties of a particular system being studied. Since here we are considering idealized HIT, the outer scale $L_L$ can be chosen arbitrarily. Once $L_L$, as well as the energy injection rate $\epsilon_{inj}$ and viscosity $\nu$, are set to provide the desired nominal $Re_{\lambda^*}$ of the pair of L/SES calculations, as described in § II B above, the SES cell size is then chosen to ensure that $\Delta x_S \lesssim \eta^*/2$ and $k_{max}\eta^* \gtrsim 6.28$. Here

$$\eta^* = \left(\frac{\nu^3}{\epsilon_{inj}}\right)^{1/4} \qquad (12)$$

is the reference Kolmogorov length scale, $\epsilon_{inj}$ is given in § II B, viscosity $\nu$ is specified for each calculation below, and $k_{max} = 2\pi/(2\Delta x_S)$ is the maximum wavenumber in the SES. This resolution has been shown in prior studies to be adequate for capturing the critical fine-scale turbulence structure[35–37].

At the same time, accuracy and stability of the L/SES crucially rely on the careful selection of a number of additional numerical parameters. These are: (i) filter scale, $\Delta$, (ii) SES domain width, $L_L$, (iii) SES sponge region size, $D_s$, (iv) LES grid resolution, $\Delta x_L$, and (v) frequency of the snapshots of the LES forcing data, $\tau_L$. We do not discuss the buffer size since, in contrast to MDP, buffer is not used in the calculations presented here.

In MDP, the effect of the sponge region size was quantified using an idealized, synthetic, 1D signal. As discussed in § II C, the purpose of the sponge region is to ensure the compatibility of the SES and LES solutions at the boundaries. In particular, the SES flow field, which contains a small-scale component, must be gradually blended with the LES solution imposed at the boundary, which consists only of the large scales. A sponge region too small would cause an abrupt change in the flow structure between the interior of the SES domain and its boundary, which would destabilize the solution. On the other hand, while a very large sponge region would benefit the solution accuracy, it would lead to a significant increase in the computational cost of the method as the governing equations are also solved in the sponge zones, while the data there cannot be used for analysis. In MDP, it was found that $D_s = \Delta/2$ provides an optimal choice, which is used in all L/SES calculations discussed below.

For all other numerical parameters, the choice of the appropriate values was discussed in MDP based on the qualitative considerations. Here we provide a more systematic assessment of all these parameters using a series of tests based on the HIT calculations.

### A. Filter scale, $\Delta$, and SES domain size, $L_S$

The central parameter, which controls the accuracy and stability of an L/SES calculation is the filter scale $\Delta$. There exist several competing considerations for the choice of $\Delta$. First, in the context of the LES, $\Delta$ must be sufficiently smaller than the LES domain size $\Delta \ll L_L$ in order to allow the development of an inertial range of scales, which are not affected by the details of the large-scale energy injection. Next, in the case of explicitly filtered LES[38–40], it is reasonable to use the same filter scale as the one used for advancing the LES calculation itself. In contrast, as discussed in MDP, in implicitly filtered LES, including implicit LES used here, $\Delta$ should be chosen in a way that would eliminate smaller scales, which are affected by the numerical grid effects or inaccuracies in the LES SGS model. In other words, even implicitly filtered LES need to be explicitly filtered for the purposes of L/SES. Therefore, the ratio $\Delta/\Delta x_L$ would depend on the details of a numerical method used, and more specifically on the spectral dependence of its numerical dissipation at small scales. Finally, in addition to these numerical considerations there also exists a physical one. Since LES typically aim to capture the inertial range in HIT but not the physical dissipation range, $\Delta$ should be comparable to the Taylor scale, $\lambda$.

On the SES side, the filter scale directly determines the choice of the SES domain size, $L_S$, since scales up to and including $\Delta$ must be explicitly resolved in the SES and thus $\Delta \leq L_S$. Therefore, it is desirable to choose $\Delta$ as small as possible to reduce the size of the SES domain, and thus the overall computational cost of the SES. On the other hand, $\Delta$, and thus $L_S$, also must be much larger than $\eta$ to allow a sufficiently large range of small scales to develop naturally with minimal impact of the SES forcing or boundary conditions. Thus in summary, the following requirements can be formulated for the choice of $\Delta$

$$L_L \gg \Delta \gg \Delta x_L; \quad \Delta \geq \lambda; \quad L_S \geq \Delta \gg \eta. \tag{13}$$

Note, as discussed above, $\Delta x_S \leq \eta$.

To make these requirements more quantitative, the effect of the choice of the filter scale is systematically examined through a series of SES, with the forcing data obtained from the same LES filtered at different $\Delta$. Parameters of the simulations are summarized in Table I. The LES was performed on a grid with size $128^3$. Kinematic viscosity both in the LES and SES was set to $\nu = 0.147$ cm$^2$/s to provide a nominal Re$_{\lambda^*} = 156$ based on eq. (5). It was also used to calculate the nominal Taylor scale $\lambda^*$ based on eq. (6), which in turn was used to set the values of $\Delta/\lambda^*$, and thus $L_S/\Delta$ in the SES given in Table I. All SES use a sponge region of size $D_s = \Delta/2$ near each boundary, as discussed above. Thus $N$ gives the number of grid cells in the active region of the SES with size $L_S$, while $N_d$ corresponds to the number of cells in the full SES domain, including the sponge zone, with size $L'_S = L_S + \Delta$.

The two main parameters varied in these tests were $\Delta$ and $L_S$. More specifically, $\Delta$ was varied between $\lambda^*$ and $4\lambda^*$, while $L_S$ was simultaneously varied between $\Delta$ and $4\Delta$, in accordance with eq. (13). Since energy injection rate in the LES is fixed, nominal $\eta^*$ (eq. 12) was set constant in all SES, namely $\eta^* = 2\Delta x_S$, as discussed above. Consequently, the ratio of the LES and SES cell sizes was also constant in all tests, $\Delta x_L/\Delta x_S = 8$, though the ratio $\Delta/\Delta x_L$ varied between 8 and 32 for different $\Delta$.

Actual values of the Taylor scale $\lambda$ relative to the actual Kolmogorov scale, $\eta$, listed in Table I were calculated in each SES as

$$\lambda = \frac{u_{rms}}{\langle (\partial u_i/\partial x_i)_S^2 \rangle^{1/2}}, \tag{14}$$

$$\eta = \left( \frac{\nu^3}{\epsilon} \right)^{1/4}. \tag{15}$$

Here, $u_{rms}$ is the actual r.m.s. velocity in the LES domain, while the velocity derivatives in eq. (14) are calculated based on the SES data, which captures the small-scale flow structure. Note that $u_{rms}$ can also be calculated in the SES, since it was shown in MDP that $u_{rms}$ in the SES is close to the actual $u_{rms}$ in a fully resolved flow. Average dissipation rate is obtained from the SES as $\epsilon = \langle 2\nu S_{ij} S_{ij} \rangle$, where $S_{ij}$ is the strain-rate tensor and $\langle . \rangle$ indicates spatial and temporal ensemble averaging. Table I also lists the corresponding Re$_\lambda = u_{rms}\lambda/\nu$, as well as the integral length scale

$$l = \frac{\pi}{2u_{rms}^2} \int_0^{k_{L,max}} \frac{E(k)}{k} dk. \tag{16}$$

Here, $E(k)$ is the TKE spectral density in the LES, $k = 2\pi/x$ is the wavenumber associated with a scale $x$, and $k_{L,max} =$

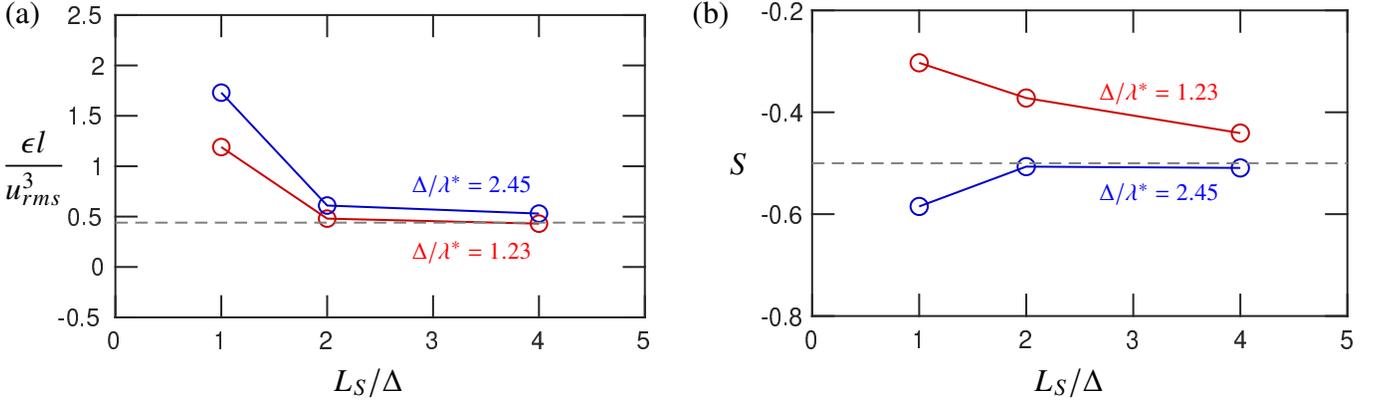

FIG. 2. (a) Normalized average dissipation and (b) skewness of the longitudinal velocity derivatives for two filter scales $\Delta/\lambda^* = 1.23$ (red) and $\Delta/\lambda^* = 2.45$ (blue) at the nominal $Re_{\lambda^*} = 156$ for the tests listed in Table I. The gray dashed line represents the theoretical estimate[30] (also see eq. 24) (a) and the reference value $S = -0.5$ in (b). Test T-$\Delta$-4-1 is not shown for clarity (see text for further details).

| Test | | $N_d$ | $N$ | $\Delta/\lambda^*$ | $L_S/\Delta$ | $\Delta x_L/\Delta x_S$ | $\Delta/\Delta x_L$ | $\eta^*/\Delta x_S$ | $Re_\lambda$ | $l/\eta$ | $\lambda/\eta$ | $\epsilon l/u_{rms}^3$ | $S$ |
|---|---|---|---|---|---|---|---|---|---|---|---|---|---|
| | LES | — | 128 | — | — | — | — | — | — | — | — | — | — |
| T-$\Delta$-1-1 | SES | 128 | 64 | 1.23 | 1 | 8 | 8 | 2 | 122 | 152 | 24 | 1.19 | -0.30 |
| T-$\Delta$-1-2 | SES | 192 | 128 | 1.23 | 2 | 8 | 8 | 2 | 181 | 109 | 30 | 0.48 | -0.37 |
| T-$\Delta$-1-4 | SES | 320 | 256 | 1.23 | 4 | 8 | 8 | 2 | 187 | 107 | 30 | 0.43 | -0.44 |
| T-$\Delta$-2-1 | SES | 256 | 128 | 2.45 | 1 | 8 | 16 | 2 | 84 | 136 | 17 | 1.73 | -0.58 |
| T-$\Delta$-2-2 | SES | 384 | 256 | 2.45 | 2 | 8 | 16 | 2 | 123 | 106 | 22 | 0.61 | -0.51 |
| T-$\Delta$-2-4 | SES | 640 | 512 | 2.45 | 4 | 8 | 16 | 2 | 131 | 105 | 23 | 0.53 | -0.51 |
| T-$\Delta$-4-1 | SES | 512 | 256 | 4.9 | 1 | 8 | 32 | 2 | 79 | 86 | 18 | 1.01 | -0.59 |

TABLE I. Parametric study of the effects of the filter width, $\Delta$, and SES domain size, $L_S$, at the nominal $Re_{\lambda^*} = 156$. In all SES, the temporal interval of the LES forcing data snapshots is $\Delta t_L \leq 0.75\tau_\Delta$, where $\tau_\Delta = \Delta/u_{rms}^*$ (eq. 21) is the characteristic filter-scale crossing time. Quantities to the left of the vertical line are simulation inputs, while quantities to the right are calculated from the simulation data. See text for the definitions of various quantities.

$2\pi/2\Delta x_L$ is the largest wavenumber associated with the LES cell size $\Delta x_L$. While the TKE spectrum in the LES does not contain all physical small scales, their contribution to $l$ is small, and the integral scale obtained using eq. (16) only from the LES, which resolves the Taylor scale, is sufficiently accurate.

Table I also lists the values of the skewness of the longitudinal velocity derivatives obtained in each SES

$$S = -\left\langle\left(\frac{\partial u_i}{\partial x_i}\right)^3\right\rangle \Big/ \left\langle\left(\frac{\partial u_j}{\partial x_j}\right)^2\right\rangle^{(3/2)}. \quad (17)$$

The results of this parametric study are illustrated in Fig. 2, which shows the values given in Table I of the normalized average dissipation (Fig. 2a) and skewness (Fig. 2b) for two filter scales $\Delta/\lambda^* = 1.23$ and $2.45$ as a function of $L_S/\Delta$. These results show that the dissipation approaches the analytical estimate[30] and converges at $L_S/\Delta \geq 2$ for both $\Delta/\lambda^* = 1.23$ and $2.45$. At the same time, skewness, which is a higher-order moment of the velocity gradients, converges for $L_S/\Delta \geq 2$ only at $\Delta/\lambda^* = 2.45$. These tests indicate, first, that the SES active region should be at least twice larger than the filter width to allow the LES forcing data to be injected into SES over some small but finite range of scales. Second, $\Delta$ should also be at least twice larger than $\lambda$ to ensure that the filter scale lies in the inertial range and the filtered forcing data from the LES is not affected by dissipation.

Finally, note that test T-$\Delta$-4-1 with the largest filter width $\Delta = 4\lambda$ (and $L_S = \Delta$) gave much less accurate results, and thus it is not shown in Fig. 2. In particular, it resulted in a much lower $Re_\lambda$ and thus much larger dissipation. It also gave much lower value of skewness. This is due to the fact that in this test $\Delta$ was only $L_L/4$ and thus it was very close to the LES domain size and the LES forcing scale. This limited range of scales, therefore, was affected by the LES forcing and was not representative of the accurate inertial range dynamics, which translated into the degraded SES solution quality. This emphasizes the importance to have $L_L \gg \Delta$ in order to generate accurate large-scale forcing data.

In summary, criteria for the choice of $\Delta$ and $L_S$ given in eq. (13) can be refined to state

$$L_L \gg \Delta \geq 2\lambda; \quad L_S \geq 2\Delta. \quad (18)$$

This, however, still leaves the LES grid cell size $\Delta x_L$ unconstrained, which we discuss next.

### B. LES grid resolution, $\Delta x_L$

In an LES, in addition to the domain size, which is typically problem-dependent, the second key parameter is the LES grid cell size, $\Delta x_L$. The choice of $\Delta x_L$ depends on the details of the LES method being used, including the numerical solver, SGS model, presence of explicit filter, etc. Various considerations to choose $\Delta x_L$ optimally are extensively described in the literature[42,43]. Ideally, LES solution must be independent of the resolution. This, however, may not be the case in practice, especially in the implicitly filtered LES. Furthermore, in the implicit (no-model) LES used here, grid resolution affects the solution by definition. Therefore, in such implicit LES approaches, when choosing the LES cell size, $\Delta x_L$ must be sufficiently smaller than the filter scale to minimize the numerical grid effects on the flow structure on scales $\geq \Delta$. Ultimately, the range of scales affected by the numerical dissipation depends on the numerical method employed, and thus it is difficult to formulate a universal prescription for $\Delta x_L$. Thus, here we examine the effect of the LES resolution on the SES for the higher-order, finite volume, Godunov-type numerical solver implemented in the code `Athena-RFX` (see § II A).

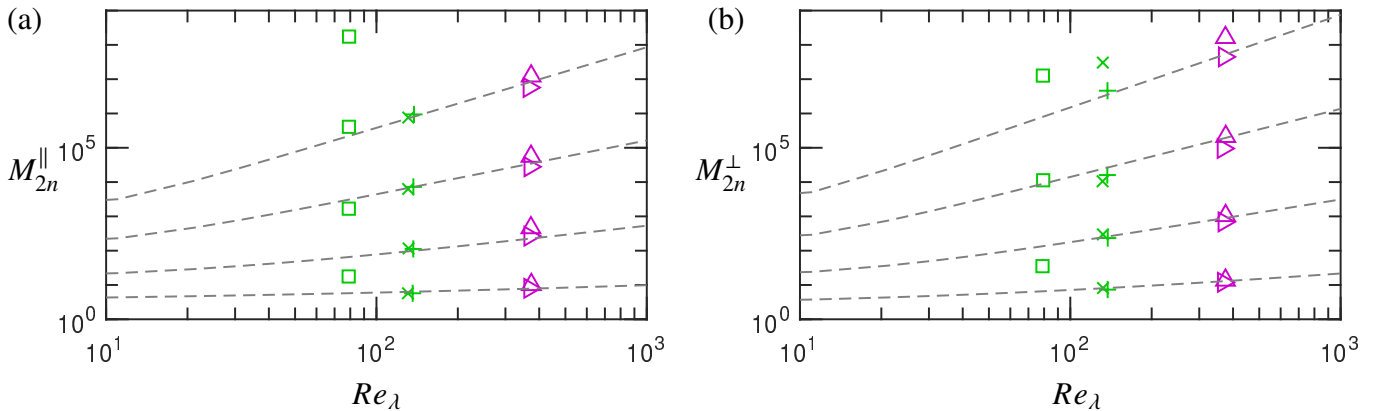

FIG. 3. Moments of the (a) longitudinal and (b) transverse velocity derivatives for tests in Table II at $Re_{\lambda^*} \approx 380$ (magenta symbols) and in Table III at $Re_{\lambda^*} \approx 150$ (green symbols). Symbols for each test are: T-$\Delta x_L$-4 $\triangleright$, T-$\Delta x_L$-8 $\triangle$, T-$\Delta t_L$-2.5 $+$, T-$\Delta t_L$-7.5 $\times$, T-$\Delta t_L$-25 $\square$. The gray dashed lines represent theoretical scalings of the moments[20,41] for $n = 2, 3, 4,$ and 5 in eq. (20) from the bottom to the top, respectively.

| Test | | $N_d$ | $N$ | $\Delta x_L/\Delta x_S$ | $\Delta/\Delta x_L$ | $L_S/\Delta$ | $\Delta/\lambda^*$ | $\eta^*/\Delta x_S$ | $Re_\lambda$ | $\lambda/\eta$ | $\epsilon l/u_{rms}^3$ | S |
|---|---|---|---|---|---|---|---|---|---|---|---|---|
| T-$\Delta x_L$-8 | LES | 512 | — | 8 | 20 | — | — | — | — | — | — | — |
| | SES | 512 | 352 | — | — | 2.20 | 2.07 | 2 | 381 | 39 | 0.45 | -0.63 |
| T-$\Delta x_L$-4 | LES | 1024 | — | 4 | 39 | — | — | — | — | — | — | — |
| | SES | 512 | 356 | — | — | 2.28 | 2.01 | 2 | 384 | 42 | 0.47 | -0.61 |

TABLE II. Parameters of the L/SES tests assessing the effect of the LES grid resolution at the nominal $Re_{\lambda^*} = 380$. In both tests, the temporal interval of the LES forcing data snapshots is $\Delta t_L \leq 0.3\tau_\Delta$, where $\tau_\Delta = \Delta/u_{rms}^*$ (eq. 21) is the characteristic filter-scale crossing time. Quantities to the left of the vertical line are simulation inputs, while quantities to the right are calculated from the simulation data. See text for the definitions of various quantities.

To examine the effect of the LES resolution, we performed two pairs of L/SES calculations at the same nominal $Re_{\lambda^*} \approx 380$, and thus the same $\nu = 0.023$ cm²/s, to probe the higher turbulent intensity regime compared to the tests in Table I. Parameters of both tests are listed in Table II. In particular, in test T-$\Delta x_L$-8, the filter scale was set to $20\Delta x_L$. This choice was motivated by the fact that in the PPM-type methods, numerical dissipation range has an extent of $\approx 10 - 20$ cells[44–46] due to the rapid drop of the numerical dissipation with scale $\propto k^{-5}$. In contrast, in test T-$\Delta x_L$-4, LES resolution was increased by a factor of two, and thus $\Delta/\Delta x_L$ was increased from 20 to 39, which reduced the ratio of the LES and SES cell sizes $\Delta x_L/\Delta x_S$ from 8 to 4. Other LES and SES parameters, were set to their optimal values as described in § III A, in particular $L_S \gtrsim 2\Delta$ and $\eta^* = 2\Delta x_S$. All other quantities listed in Table II were defined in § III A above. Taylor microscale in both SES calculated using eq. (14) was $\approx 2$ times smaller than $\Delta$ in agreement with the requirement in eq. (18), and the resulting actual $Re_\lambda$ was close to the target nominal value.

Table II shows that values of the normalized average dissipation rate and skewness are very close for the two LES resolutions. To verify this observed solution convergence, we consider another important and very sensitive set of metrics of the quality of the turbulent-flow solution, namely the higher-order moments of velocity derivatives. Specifically, the following even-order moments of the longitudinal and transverse velocity derivatives are defined as

$$M_{2n}^{\parallel} = \left\langle \left(\frac{\partial u_i}{\partial x_i}\right)^{2n} \right\rangle / \left\langle \left(\frac{\partial u_i}{\partial x_i}\right)^2 \right\rangle^n, \quad (19a)$$

$$M_{2n}^{\perp} = \left\langle \left(\frac{\partial u_i}{\partial x_j}\right)^{2n} \right\rangle / \left\langle \left(\frac{\partial u_i}{\partial x_j}\right)^2 \right\rangle^n. \quad (19b)$$

While the pioneering work on the moments dates back to Kolmogorov[47], recently substantial progress has been made towards generalizing the analytical formulae for the scaling of the 4th- to the 10th-order moments. In particular, it has been suggested that both $M_{2n}^{\parallel}$ and $M_{2n}^{\perp}$ have a power-law dependence on $Re_\lambda$, when $Re_\lambda$ is sufficiently large.[19,20] A general form for the scaling of the higher-order moments was proposed[41]

$$M_{2n} \sim a + (Re_\lambda/b)^\beta. \qquad (20)$$

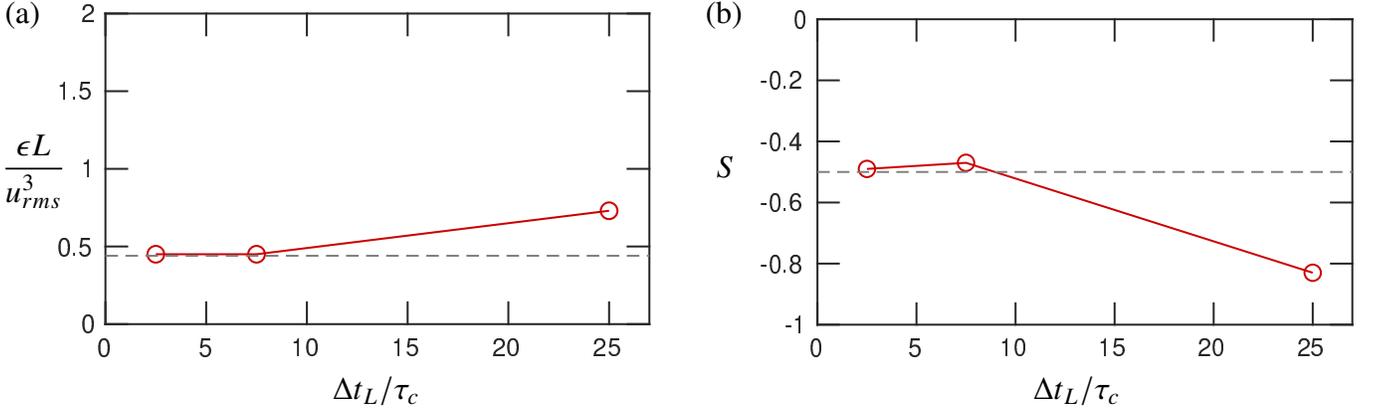

FIG. 4. (a) Normalized average dissipation and (b) skewness of the longitudinal velocity derivatives as a function of the time interval between the LES forcing data snapshots, $\Delta t_L$, normalized by the characteristic cell-crossing time, $\tau_c$. The gray dashed line represents the theoretical estimate[30] (also see eq. 24) in (a) and the reference value of $-0.5$ in (b).

| Test | | $N_d$ | $N$ | $\Delta t_L/\tau_c$ | $\Delta t_L/\tau_\Delta$ | $L_S/\Delta$ | $\Delta x_L/\Delta x_S$ | $\Delta/\Delta x_L$ | $\Delta/\lambda^*$ | $\eta^*/\Delta x_S$ | $Re_\lambda$ | $l/\eta$ | $\lambda/\eta$ | $\epsilon L/u_{rms}^3$ | $S$ |
|---|---|---|---|---|---|---|---|---|---|---|---|---|---|---|---|
| | LES | — | 128 | — | — | — | — | — | — | — | | | | | |
| T-$\Delta t_L$-2.5 | SES | 384 | 256 | 2.5 | 0.16 | 2 | 8 | 16 | 2.45 | 2 | 137 | 90 | 23 | 0.45 | -0.49 |
| T-$\Delta t_L$-7.5 | SES | 384 | 256 | 7.5 | 0.47 | 2 | 8 | 16 | 2.45 | 2 | 132 | 91 | 25 | 0.45 | -0.47 |
| T-$\Delta t_L$-25 | SES | 384 | 256 | 25 | 1.56 | 2 | 8 | 16 | 2.45 | 2 | 79 | 63 | 17 | 0.73 | -0.83 |

TABLE III. Parametric study of the time interval of LES data snapshots, $\tau_L$, at the nominal $Re_{\lambda^*} = 156$. Quantities to the left of the vertical line are simulation inputs, while quantities to the right are calculated from the simulation data. See text for the definitions of various quantities.

The values of the scaling exponent $\beta$ for $n = 2$ to 5 are 0.36, 0.91, 1.58, and 2.35, respectively, for the longitudinal moments, and 0.62, 1.28, 1.99, and 2.7, respectively, for the transverse moments.[41] The values of $a$ and $b$ do not affect the $Re_\lambda$-scaling of the moments, and thus we do not include them here. Such values, along with the discussion of the associated uncertainties, can be found in Refs.[20,41].

Values of $M_{2n}^\parallel$ and $M_{2n}^\perp$ for tests T-$\Delta x_L$-8 and T-$\Delta x_L$-4 are shown in Fig. 3 (magenta triangles). Data from these two SES cases are very similar and are consistent with the analytical predictions from the literature (dashed gray lines).

Results in Table II and Fig. 3 show that LES resolution of $\Delta/\Delta x_L \approx 20$ provides a converged solution in close agreement with the theoretical predictions for various turbulence metrics, including higher-order moments. Furthermore, this demonstrates that L/SES provides an accurate turbulent flow field even when the SES large-scale forcing data is obtained from a coarse implicit LES with $\Delta x_L = 4\eta$, which does not resolve the dissipation range. In all L/SES simulations described below, we adopt this (or finer) LES resolution.

### C. Time interval of the LES snapshots for forcing, $\Delta t_L$

The LES data for forcing the SES is available at discrete time intervals $\Delta t_L$. Similar to other L/SES parameters, there exist competing requirements for setting the optimal value of $\Delta t_L$. On one hand, $\Delta t_L$ should be as small as possible to minimize the errors associated with temporal interpolation of the forcing flow field to the current SES time step. Note that $\Delta t_L$ cannot be smaller than the LES time steps, which are by definition larger than the time steps in the SES due to the difference in resolution. On the other hand, larger values of $\Delta t_L$ are desirable to minimize the cost of filtering and spatial interpolation of the LES forcing data, which is significant relative to the computational cost of advancing the overall LES solution.

It was suggested in MDP that an optimal choice for $\Delta t_L$ is based on the characteristic filter-scale crossing time, i.e.,

$$\Delta t_L \leq \frac{\Delta}{u_{rms}} \equiv \tau_\Delta. \qquad (21)$$

On this timescale, the flow changes primarily on scales $\leq \Delta$, which are removed by filtering. As a result, the large-scale flow structures, which are transferred to SES, do not evolve significantly and thus they can be interpolated with a minimal loss of accuracy. If $\Delta/\Delta x_L \sim 16$, as discussed in § III B, then $\Delta t_L/\tau_c \lesssim 16$, where $\tau_c \equiv \Delta x_L/u_{rms}$ is the characteristic cell-crossing time by the flow.

To assess the accuracy of this prescription and determine the optimal value of $\Delta t_L$, three SES were performed, which were forced with the data extracted from the same LES at three different time intervals. In particular, $\Delta t_L/\tau_\Delta$ was varied between $\approx 1/3$ and 3, and correspondingly $\Delta t_L/\tau_c$ varied between 2.5 and 25. LES calculation was the same as the one used in tests in Table I with the nominal $Re_{\lambda^*} \approx 150$. All SES parameters shown in Table III, namely $L_S$, $\Delta$, $\Delta x_L$, and $\Delta x_S$ were set to their optimal values as described above.

Resulting values of the normalized average dissipation and skewness of the longitudinal velocity derivatives are listed in Table III and are shown in Fig. 4. Both quantities approach their reference values and saturate at $\Delta t_L/\tau_c \lesssim 7.5$, or $\Delta t_L/\tau_\Delta \lesssim 1$.

Furthermore, values of the higher-order moments of the longitudinal and transverse velocity derivatives for these three SES tests are shown in Fig. 3 (green symbols) along with the corresponding theoretical scaling[41]. For the largest value of $\Delta t_L/\tau_c = 25$ (green square), the SES significantly underestimates $Re_\lambda$ and overestimates the values of the momends of all orders. At the same time, for $\Delta t_L/\tau_c \lesssim 7.5$, or equivalently for $\Delta t_L < \tau_\Delta$, the values of all moments are virtually the same and they are in close agreement with the proposed theoretical scaling. The only exception is $M_{10}^\perp$, which is somewhat overestimated for $\Delta t_L/\tau_c = 7.5$.

This shows that the original criterion in eq. (21) proposed in MDP indeed provides the optimal choice for $\Delta t_L$. It ensures high accuracy of the SES flow solution, while at the same time allowing filtering and interpolation of the LES data to be performed every $\sim 100$ timesteps for the low-Mach-number turbulence considered here (cf. eq. 12 in MDP) resulting in a negligible relative cost of these operations in LES.

### D. Summary of the L/SES parameters

Based on the parametric studies discussed above, the optimal L/SES parameters, which are used in the L/SES discussed below, can be summarized as follows:

- Filter width: $\Delta \geq 2\lambda$;
- LES resolution: $\Delta x_L \leq \Delta/16$;
- LES data time interval: $\Delta t_L \leq \Delta/u_{rms}$;
- SES domain size (active region): $L_S \geq 2\Delta$;
- SES sponge region size: $D_S \geq \Delta/2$;
- SES resolution: $\Delta x_S \leq \eta/2$.

It can be seen that all L/SES parameters are determined by the two turbulence scales: inner (dissipative) scale, $\eta$, and Taylor scale, $\lambda$. Finally, as will be further shown below, the accuracy of the L/SES improves for larger $Re_\lambda$ as the separation between all scales, in particular $L_L$ and $\lambda$, as well as between $\lambda$ (and thus $L_S$ and $\Delta$) and $\eta$, increases.

## IV. RESULTS

To investigate the accuracy of the L/SES method for a range of turbulent conditions, we performed several L/SES for the Reynolds numbers in the range $Re_\lambda \approx 100-600$. All LES use the same domain size $L_L = 0.45$ cm and the same energy injection rate $\epsilon_{inj} = 5.29 \times 10^9$ erg/g·s, as in the tests described above. Increasing $Re_\lambda$ is achieved by decreasing viscosity. Values of $\nu$ in each calculation are listed in Table IV, along with other simulation parameters, which were chosen in accordance with the criteria outlined in § III D. In order to minimize the compressibility effects in the turbulent flow field, all simulations were performed at the same nominal turbulent Mach number $M_t = u_{rms}^*/c = 0.05$, where $c$ is the speed of sound. The extension of the L/SES method to compressible flows is the subject of future work. Finally, to allow direct quantitative comparison, for the two lowest Reynolds numbers $Re_\lambda \approx 100$ and 150, the DNS counterparts were also performed with the same $\epsilon_{inj}$, $\nu$, and domain size as in the LES, and with the same resolution $\Delta x_S$ as in the SES.

All simulations were performed over a period of $\tau_{sim}$ listed in Table IV. Each LES calculation was allowed to reach fully developed turbulence for at least $6\tau_{ed}$, after which time the LES flow field data were collected at discrete time intervals $\Delta t_L$, filtered, interpolated, and used to force the companion SES, as described in § II B. The first one of these LES data snapshots served as the initial condition for the SES. Once an SES was initialized, its flow field again was allowed to equilibrate with the LES over the first $2\tau_{ed}$ before data analysis would begin. All statistics reported here were time-averaged over the remaining duration of the SES. Finally, $k_{max}\eta^*$ for each case was set to 6.28, while the $k_{max}\eta$ listed in Table IV is based on the actual $\eta$ obtained from the small-scale dissipation in the SES.

### A. Global turbulence characteristics

Figure 5(a) shows spectra of the specific TKE in DNS and L/SES at $Re_\lambda \approx 100$. Spectra are normalized by the total TKE, $u_{rms}^2/2 = u_i u_i/2$, and by the integral length scale, $l$. TKE spectra in the non-periodic SES domain are calculated following the procedure described in MDP, and they are shown only up to the filter scale, since larger scales are affected by forcing. The figure shows close agreement between the LES and DNS at the energy-containing large scales. This confirms the main premise of the L/SES approach, namely that large-scale motions in the LES are accurately represented due to the forward energy cascade. LES and DNS spectra start to deviate at scales close to the filter scale $\Delta$ (vertical dotted line in Fig. 5a) and smaller, however these scales in the LES are removed by filtering prior to their injection into the SES.

The SES spectrum in Fig. 5a, which represents the dissipation range on scales $\leq \Delta$, virtually coincides with the DNS spectrum with the exception of the smallest scales $\leq \eta$, which are close to the grid resolution and thus are affected by numerical dissipation. Finally, the scaled SES spectra for the entire range of $Re_\lambda \approx 100-600$ considered here are shown in Fig. 5(b). It can be seen that that the energy cascade extends to higher wavenumbers with increasing $Re_\lambda$ as a result of decreasing viscosity.

In addition to spectra, L/SES must be able to recover proper $Re_\lambda$ dependence of various characteristic scales of turbulence, namely integral (outer), $l$, dissipative (inner), $\eta$, and Taylor scale, $\lambda$, in particular[48]

$$l/\eta \sim Re_\lambda^{3/2}, \quad \lambda/\eta \sim Re_\lambda^{1/2}. \tag{22}$$

These scalings reflect the growing scale separation in a turbulent flow with increasing $Re_\lambda$. Similarly, TKE, or equivalently

| Case | | $N_d$ | $N$ | $\nu$ (cm²/s) | $Re_\lambda$ | $\Delta x_L/\Delta x_S$ | $\Delta/\lambda$ | $\Delta/\Delta x_L$ | $L_S/\Delta$ | $\Delta t_L/\tau_\Delta$ | $\eta/\Delta x_S$ | $k_{max}\eta$ | $\tau_{sim}/\tau_{ed}$ |
|---|---|---|---|---|---|---|---|---|---|---|---|---|---|
| | DNS | 512 | — | 0.370 | 88 | — | — | — | — | — | 2.0 | 6.22 | 10 |
| Re-100 | LES | 128 | — | " | — | — | — | — | — | — | — | — | 25 |
| | SES | 256 | 176 | " | 108 | 4 | 2.08 | 20 | 2.2 | 0.15 | 1.9 | 5.85 | 20 |
| | DNS | 1024 | — | 0.147 | 147 | — | — | — | — | — | 2.0 | 6.29 | 10 |
| Re-150 | LES | 256 | — | " | — | — | — | — | — | — | — | — | 20 |
| | SES | 320 | 220 | " | 153 | 4 | 2.03 | 25 | 2.2 | 0.24 | 1.8 | 5.78 | 15 |
| Re-220 | LES | 512 | — | 0.058 | — | — | — | — | — | — | — | — | 20 |
| | SES | 384 | 260 | " | 215 | 4 | 2.05 | 31 | 2.1 | 0.39 | 1.8 | 5.72 | 15 |
| Re-400 | LES | 512 | — | 0.023 | — | — | — | — | — | — | — | — | 15 |
| | SES | 512 | 352 | " | 375 | 8 | 2.07 | 20 | 2.2 | 0.61 | 1.9 | 5.91 | 11 |
| Re-600 | LES | 1024 | — | 0.0092 | — | — | — | — | — | — | — | — | 15 |
| | SES | 640 | 440 | " | 575 | 8 | 2.05 | 25 | 2.2 | 0.49 | 1.8 | 5.66 | 10 |

TABLE IV. Summary of the DNS and L/SES performed. All parameters have the same meaning as in Tables I-III and as defined in §§ II and III.

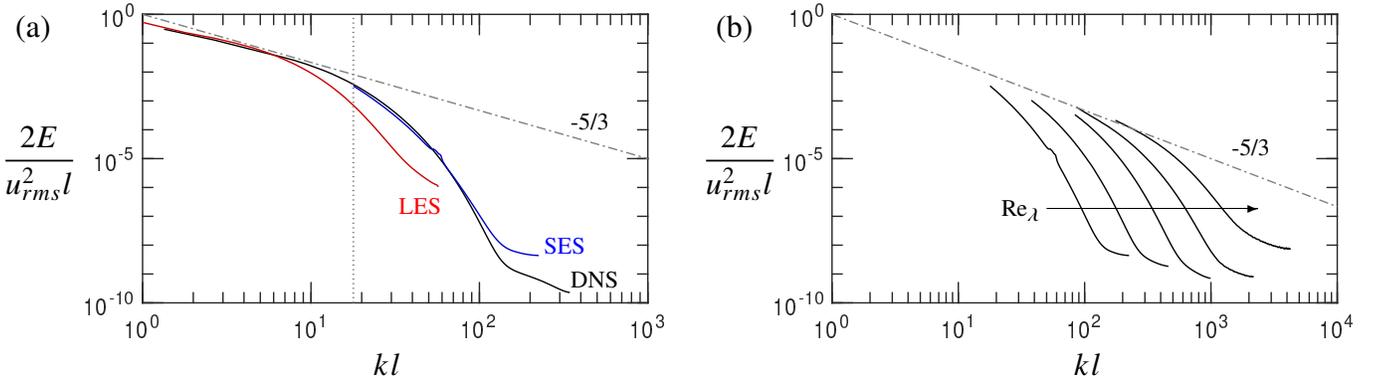

FIG. 5. Normalized turbulent kinetic energy spectra. (a) Spectra for case Re-100 at $Re_\lambda \approx 100$ from DNS (black), LES (red), and SES (blue). The vertical dotted line represents the filter width, $\Delta$. (b) Spectra from SES for cases Re-100, Re-150, Re-220, Re-400, and Re-600 from left to right, respectively. The gray dash-dotted lines in both panels represent the $k^{-5/3}$ slope for reference.

$u_{rms}$, must also follow the $Re_\lambda$ scaling

$$u_{rms}/u_\eta \sim Re_\lambda^{1/2}, \qquad (23)$$

where $u_\eta = (\nu\epsilon)^{1/4}$ is the Kolmogorov velocity scale.

Ensemble- (time- and space-) averaged values of these ratios obtained in L/SES for $Re_\lambda \approx 100 - 600$ are shown in Fig. 6. For comparison, the values obtained in DNS for the two lowest $Re_\lambda$ are also shown with red circles. All quantities agree well with the theoretical power laws of $Re_\lambda$. In particular, note that the agreement with eqs. (22) and (23) becomes progressively better with increasing $Re_\lambda$, with some discrepancy present for $l/\eta$ at the lowest $Re_\lambda \approx 100 - 150$. This discrepancy arises mainly due to the fact that $l$ is calculated in eq. (16) based on the TKE spectrum obtained only from LES. At lower $Re_\lambda$, and therefore in smaller LES domains, small scales affected by the unphysical numerical dissipation (cf. Fig. 5) have a greater relative contribution to the integral of the $k$-weighted spectral energy density in eq. (16), which in turn introduces error in the resulting value of $l$. As $Re_\lambda$ increases, the extent of the inertial range in LES increases, thus increasing turbulent scale separation and improving the accuracy of the L/SES solution. The importance of having the outer turbulent scale sufficiently larger than the filter scale in order to achieve high L/SES solution accuracy was emphasized in § III A above. This is not a limitation of the method since L/SES is primarily intended for high-$Re_\lambda$ flows, in which DNS become prohibitively costly. Finally, since the main goal of L/SES is to recover the flow structure on smaller scales, errors in large-scale quantities, namely integral scale $l$, are less critical. Smaller-scale quantities, such as $\lambda$ and $\eta$, are properly recovered even at lower $Re_\lambda$, as shown in Fig. 6.

Another important question related to the coupling of large and small scales in L/SES concerns the ability of the L/SES to capture the dissipative anomaly properly. It was previously argued[13,17,30,49–52] that at high $Re_\lambda$, normalized turbulent dissipation $\epsilon l/u_{rms}^3$ approaches a constant value close to 0.5 representative of a universal asymptotic state[50]. In particular, a theoretical expression was suggested[30,50]

$$\epsilon l/u_{rms}^3 = A(1 + \sqrt{1 + (B/Re_\lambda)^2}), \qquad (24)$$

where $A = 0.2$ and $B = 92$. Note that other theoretical prescriptions for the dissipation can also be found in the literature[53–55]. Figure 7 shows the normalized dissipation from both DNS and L/SES listed in Table IV, along with the representative data from the literature[17]. Again, close agreement is observed with the analytical expression for a wide range of $Re_\lambda$, with the normalized dissipation from the present work reaching an asymptotic value $\approx 0.5$ with increasing $Re_\lambda$. Somewhat suppressed normalized dissipation, which can be seen at lower $Re_\lambda \approx 100 - 150$, is again the result of the lower values of $l$ discussed above (cf. Fig. 6a).

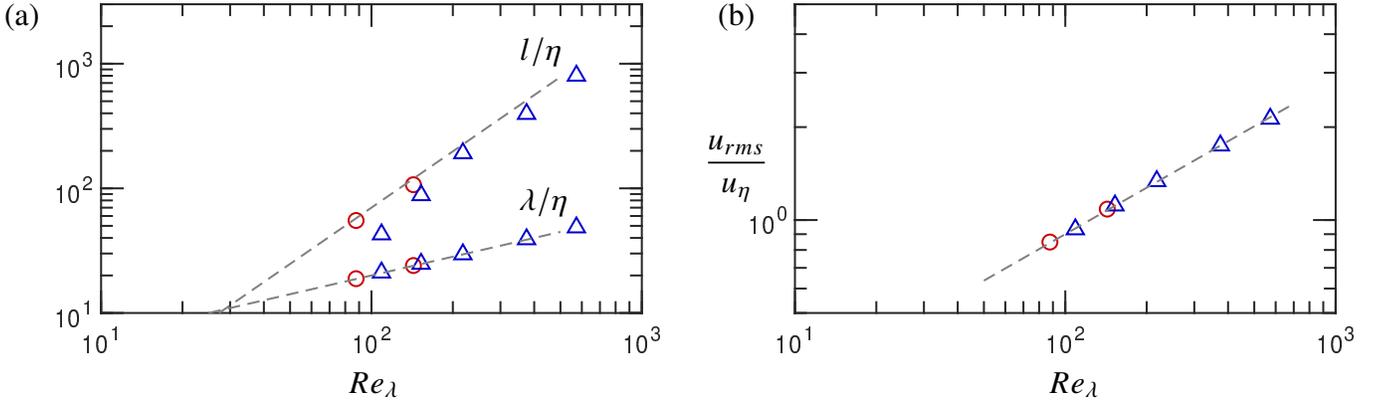

FIG. 6. (a) Scaling of the integral length scale, $l$, and the Taylor micro-scale, $\lambda$, normalized by the Kolmogorov length scale, $\eta$. The dashed lines represent theoretical scaling (eq. 22). (b) Scaling of the r.m.s. velocity, $u_{rms}$, normalized by the Kolmogorov velocity scale, $u_\eta$. The dashed line represents theoretical scaling (eq. 23). In both panels, shown are values for DNS ($\bigcirc$) and L/SES ($\triangle$).

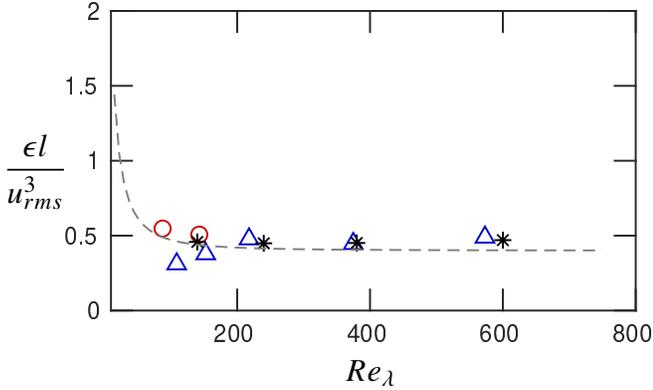

FIG. 7. Normalized dissipation of the turbulent kinetic energy in DNS ($\bigcirc$) and L/SES ($\triangle$). Also shown are the values from the literature[17] ($*$). The dashed line represents eq. (24).

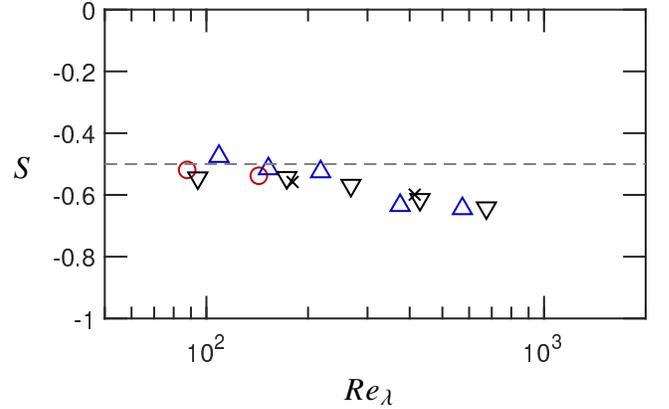

FIG. 8. Skewness of the longitudinal velocity derivatives (eq. 17) from DNS ($\bigcirc$) and L/SES ($\triangle$). Also shown are the reference data from the literature ($\times$)[56] and ($\triangledown$)[18] (black symbols). Horizontal dashed line represents the reference value of $-0.5$.

### B. Moments of the velocity derivatives

Next we describe the small-scale turbulence characteristics for $Re_\lambda \approx 100 - 600$. Figure 8 shows ensemble-averaged skewness of the longitudinal velocity derivatives (eq. 17) in the L/SES and DNS listed in Table IV, along with the reference data from the literature. At the range of $Re_\lambda$ values considered in this study, the skewness obtained in L/SES remains close to $-0.5$, and it is in agreement with our DNS results. L/SES results also agree with the published skewness values from DNS[18,56], which are shown in Fig. 8 with black symbols.

Moments of the longitudinal and transverse velocity derivatives from the 4$^{th}$ to the 10$^{th}$ order (eqs. 19) are shown in Fig. 9.

Both the longitudinal and transverse moments are accurately captured by the L/SES when compared to the analytical solutions, even for the 10$^{th}$-order moments at $Re_\lambda \approx 400 - 600$. In agreement with the literature, our results also show that the transverse derivatives increase faster with Reynolds number than the longitudinal derivatives. It was suggested that this observation can be explained by the possible stronger intermittency in the transverse gradients than the longitudinal.[57]

### C. PDF of the velocity derivatives

Higher-order moments of the velocity increments provide information about the overall degree of intermittency in the turbulent flow, with larger values of $n$ emphasizing the contribution of more extreme events. At the same time, complete information about the statistics of the velocity derivatives is provided by their probability density functions (PDF). In particular, here we consider the PDF of enstrophy, $\Omega \equiv \omega_i \omega_i / 2$, where $\omega_i$ is the vorticity, and dissipation rate, $\epsilon \equiv 2\nu S_{ij} S_{ij}$, where $S_{ij}$ again is the strain rate tensor. Figure 10 shows the PDF of $\Omega'/\langle\Omega\rangle$ and $\epsilon'/\langle\epsilon\rangle$, where superscript ′ represents the fluctuating values and $\langle.\rangle$ represents the ensemble average.

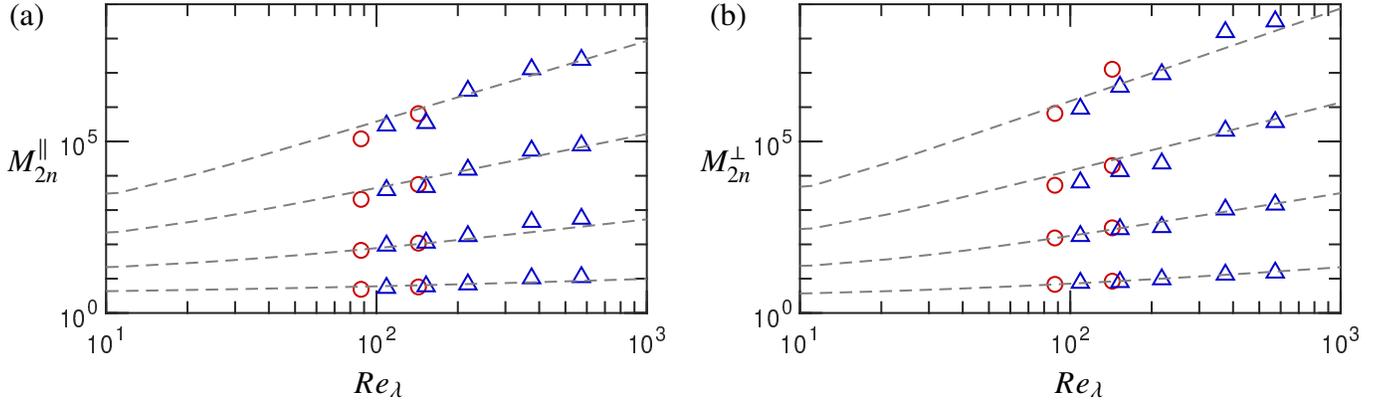

FIG. 9. Moments of the (a) longitudinal and (b) transverse velocity derivatives from DNS (○) and L/SES (△). The gray dashed lines represent theoretical scalings based on eq. (20) from Khurshid, Donzis, and Sreenivasan [41] for $n = 2, 3, 4$, and 5 from the bottom to the top, respectively.

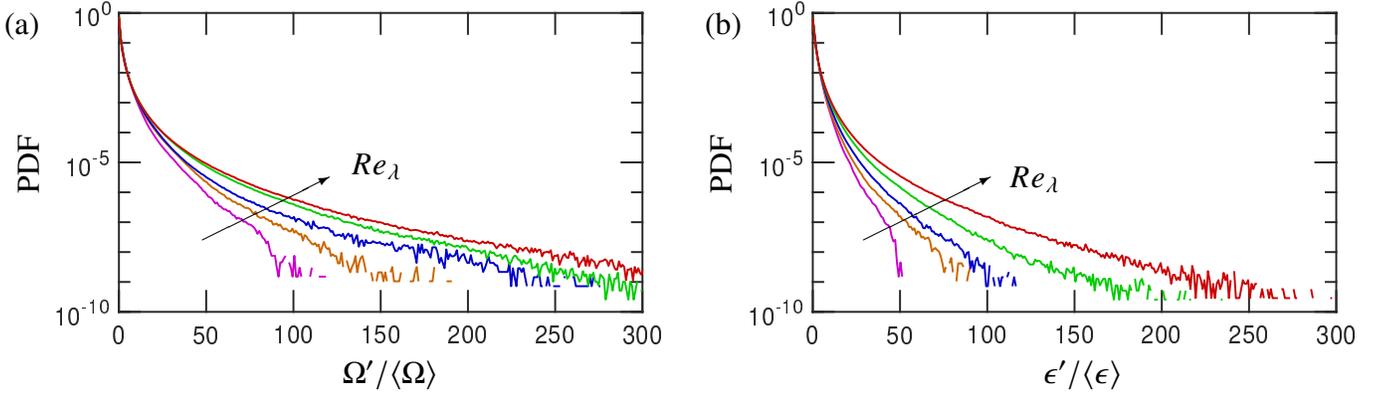

FIG. 10. The PDF of (a) enstrophy and (b) dissipation at $Re_\lambda \approx 100$ (magenta); $\approx 150$ (brown); $\approx 220$ (blue); $\approx 400$ (green); $\approx 600$ (red) from L/SES.

| $Re_\lambda$ | $b_\Omega$ | $b_\epsilon$ |
|---|---|---|
| 140 | 6.44 | 8.57 |
| 240 | 6.12 | 7.91 |
| 390 | 5.47 | 6.76 |
| 650 | 5.32 | 6.65 |

TABLE V. Best-fit values of the coefficient $b_q$ in eq. (25). [16]

With increasing $Re_\lambda$, turbulent flows become more intermittent, resulting in wider PDF with longer tails. Moreover, PDF of enstrophy are more intermittent than those of dissipation. In particular, note that PDF of $\epsilon'/\langle\epsilon\rangle$ reaches 300 and PDF of $\Omega'/\langle\Omega\rangle$ reaches 500 in the $Re_\lambda \approx 600$ case (Fig. 11e). Both trends are in agreement with the previous studies. [16,58] To make this statement more quantitative, next we compare the L/SES results with the DNS data and the analytical fits to the PDF published in the literature.

Donzis, Yeung, and Sreenivasan [16] suggested that the PDF of both enstrophy and dissipation can be approximated with a

stretched exponential function of the form

$$\text{PDF}(q) \sim \exp\left[b_q q^{0.25}\right], \qquad (25)$$

where $q$ is either $\Omega'/\langle\Omega\rangle$ or $\epsilon'/\langle\epsilon\rangle$. Such stretched exponential fits provide a more quantitative comparison of the PDF obtained here with the prior results in the literature. In particular, the best-fit coefficients suggested by Donzis, Yeung, and Sreenivasan [16] for the range of $5 < q < 100$ are listed in Table V. The values of $Re_\lambda$ for those coefficients in the table are slightly different from the present L/SES, but they are reasonably close to be used as a reference.

Figure 11 shows the PDF of $\Omega'/\langle\Omega\rangle$ and $\epsilon'/\langle\epsilon\rangle$ individually for each $Re_\lambda$ case. Also shown are the stretched exponential fits based on eq. (25) with the coefficients from Table V as black dashed and dash-dotted lines for both enstrophy and dissipation, respectively, at the closest $Re_\lambda$. Finally, for comparison, at the two lowest $Re_\lambda$, the PDF based on the DNS are shown with the red and blue dash-dotted lines. Note that at $Re_\lambda \approx 150$, the DNS PDF agree very closely with the stretched exponentials, confirming that such stretched exponential fits indeed can be viewed as a close representation of the accurate flow structure.

At the lowest $Re_\lambda \approx 100 - 150$, L/SES PDF for both $\epsilon$ and $\Omega$ do not reach the same extreme values as the DNS with the tails of the PDFs being suppressed. Furthermore, while the PDF of $\epsilon$ follows the DNS PDF quite closely at lower values of $\epsilon'/\langle\epsilon\rangle$, the enstrophy PDF is below both the DNS PDF and the exponential fits reflecting lower intermittency in the L/SES. This somewhat degraded solution accuracy at lower $Re_\lambda$ was discussed in § IV A above in the context of other quantities, including integral scale and normalized dissipation. It is the result of the insufficient scale separation both in the LES between $L_L$ and $\Delta$ and in the SES between $L_S$ and $\eta$, which affects the development of the proper turbulent inter-scale dynamics.

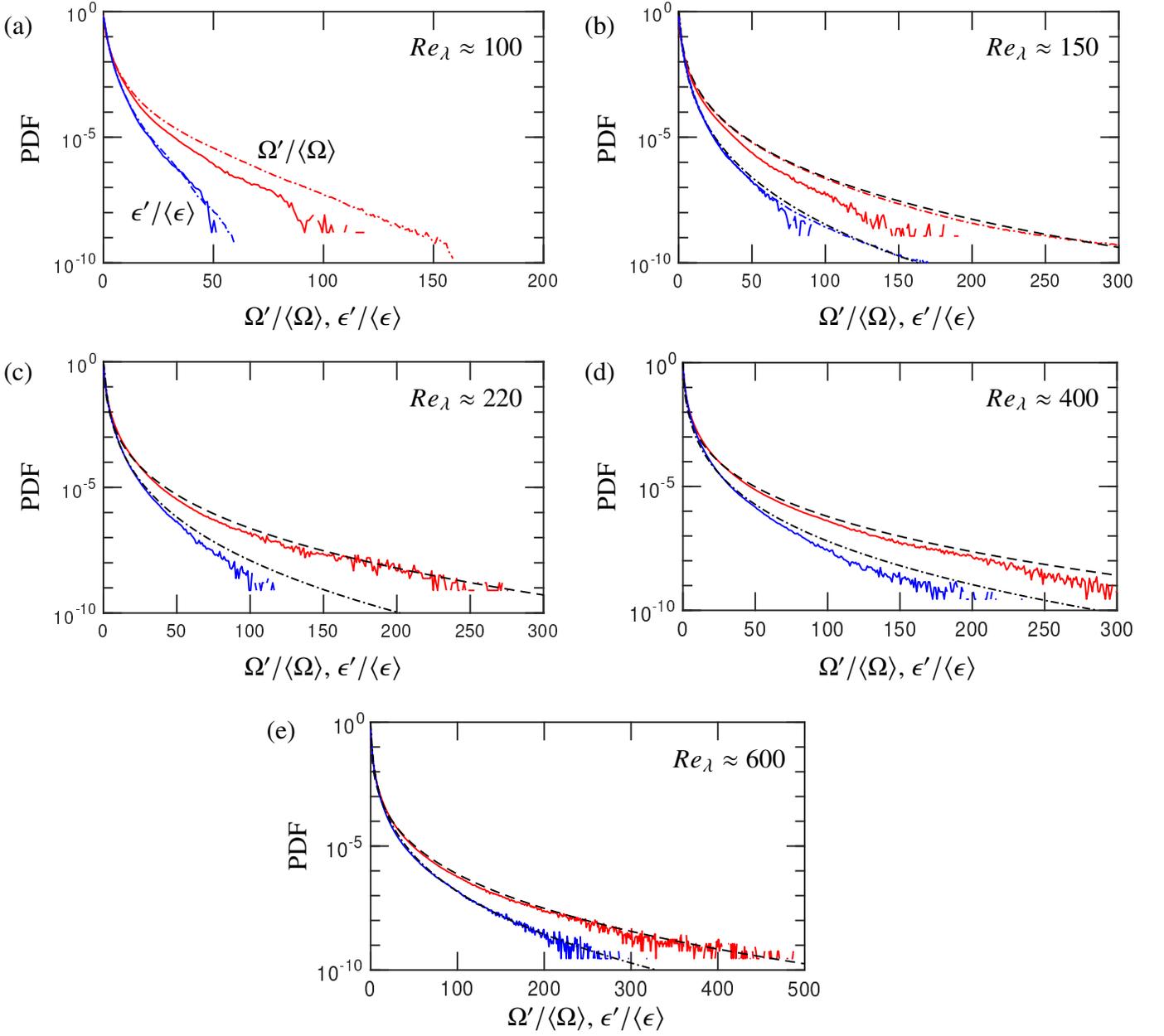

FIG. 11. The PDF of enstrophy (solid red) and dissipation (solid blue) in L/SES at (a) $Re_\lambda \approx 100$, (b) $Re_\lambda \approx 150$, (c) $Re_\lambda \approx 220$, (d) $Re_\lambda \approx 400$, and (e) $Re_\lambda \approx 600$. The red and blue dash-dotted lines in panels (a) and (b) are from DNS. The black dashed and dash-dotted lines are the stretched exponential fits (eq. 25) for enstrophy and dissipation, respectively.[16]

For higher $Re_\lambda$, both PDF extend to progressively larger values. The overall agreement with the exponential fits improves, especially for $\Omega$, and at $Re_\lambda \approx 600$, L/SES PDF trace the exponential fits closely over the entire range. The ability of the L/SES to capture such subtle turbulence property as intermittency, which reflects its complex non-linear dynamics, at high $Re_\lambda$, at which the DNS cost would become considerable, is one of the main strengths of the proposed method. Finally, we note that all PDF were confirmed to be converged.

## V. COMPUTATIONAL COST OF L/SES

Results discussed above show that the L/SES can recover the turbulence structure with an accuracy similar to the DNS performed at the same $Re_\lambda$ and resolution. Furthermore, solution accuracy, or more specifically, agreement with the theoretical predictions, improves for higher $Re_\lambda$, at which DNS become progressively more expensive. This leads to the central question of the computational efficiency, or cost savings, of the L/SES method compared to the DNS. The L/SES is more algorithmically complex, and it also carries additional computational cost per cell primarily associated with the filtering operation. Therefore, without a considerable overall increase in the computational efficiency, L/SES would not present a practical alternative to DNS.

The computational efficiency of the method exploits the difference in scaling between the energy containing large scales $\sim l \approx L_L$ and the intermediate Taylor scale $\lambda \lesssim L_S$ relative to the dissipative scales $\sim \eta \approx 2\Delta x_S$, as shown in Fig. 6a (also see eq. 22). An LES targets scales in the range between $\approx l$ and $\lambda$, so that the number of grid points in each direction is

$$N_L = \frac{L_L}{\Delta x_L} \approx \frac{l}{\lambda} \sim \frac{a_l}{a_\lambda} Re_\lambda. \quad (26)$$

A companion SES targets scales from $\approx \lambda$ down to the Kolmogorov scale $\eta$. Thus, the SES domain contains

$$N_S = \frac{L_S}{\Delta x_S} \approx \frac{\lambda}{\eta} \sim a_\lambda Re_\lambda^{1/2} \quad (27)$$

cells in each direction (see column $N_d$ in Table IV for the values of $N_L$ and $N_S$ in each case). Here, $a_l$ and $a_\lambda$ are the proportionality factors discussed next, and we use the criteria for $L_L$, $L_S$, $\Delta x_L$, and $\Delta x_S$ summarized in § III D.

In contrast, the grid size along each direction in a DNS, which must capture the full range of turbulent scales from the inner to the outer one, scales as

$$N_D = \frac{L_L}{\Delta x_S} = N_L \frac{\Delta x_L}{\Delta x_S} \approx \frac{l}{\eta} \sim a_l Re_\lambda^{3/2}. \quad (28)$$

As a result, the ratio of the number of cells in an LES and SES pair of calculations, relative to an equivalent DNS, scales as

$$\text{Cell ratio} \equiv \frac{N_L^3 + N_S^3}{N_D^3} \sim a_\lambda^{-3} Re_\lambda^{-3/2} + \left(\frac{a_\lambda}{a_l}\right)^3 Re_\lambda^{-3}. \quad (29)$$

Since $l/\eta$ approaches $\lambda/\eta$ at sufficiently low $Re_\lambda$ (see Fig. 6a), the scaling coefficient $a_l$ is much smaller than $a_\lambda$. In particular, $a_\lambda \sim O(1)$, while $a_l \sim O(10^{-2})$. As a result, the coefficient in front of the first term in eq. (29) is $O(1)$, while the coefficient in front of the second term is $O(10^6)$, and thus the second term dominates. This shows that the total number of cells in an L/SES drops very rapidly with $Re_\lambda$ relative to a DNS, unless $Re_\lambda$ becomes extremely large causing the first term to become more dominant. Yet even at such large $Re_\lambda$, L/SES cost relative to a DNS would continue to decrease with $Re_\lambda$, albeit more slowly.

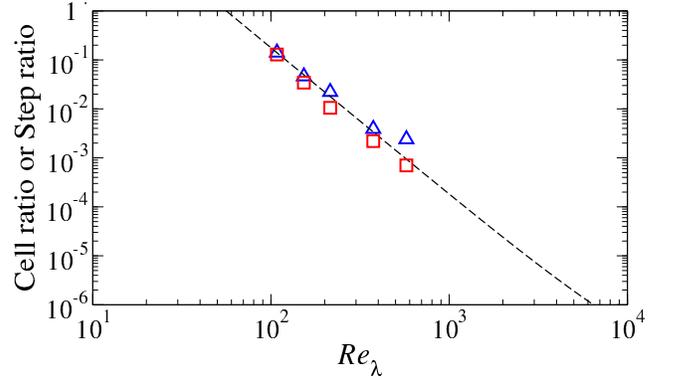

FIG. 12. Scaling of the computational cost of an L/SES relative to an equivalent DNS with $Re_\lambda$. Shown are the cell ratio (eq. 29, △) and cell step ratio (eq. 30, □). Black dashed line represents the $\sim Re_\lambda^{-3}$ scaling (see eqs. 29 and 30).

| Case | $N_L$ | $N_S$ | $N_D$ | Cell ratio | Step ratio |
|------|-------|-------|-------|-----------|-----------|
| Re-100 | 128 | 256 | 512 | 14.1% | 12.9% |
| Re-150 | 256 | 320 | 1,024 | 4.61% | 3.44% |
| Re-220 | 512 | 384 | 2,048 | 2.22% | 1.05% |
| Re-400 | 512 | 512 | 4,096 | 0.39% | 0.22% |
| Re-600 | 1,024 | 640 | 8,192 | 0.24% | 0.07% |

TABLE VI. Grid sizes of the LES, SES, and equivalent DNS, along with the corresponding cell ratio (eq. 29) and cell step ratio (eq. 30), for the five $Re_\lambda$ cases considered in § IV.

This rapid drop in the relative cost of an L/SES with $Re_\lambda$ is illustrated in Fig. 12, which shows with blue triangles the cell ratio for all cases considered in § IV (also listed in Table VI). For comparison, the $Re_\lambda^{-3}$ scaling given in eq. (29) is also shown with a dashed black line. Cell ratio becomes less than one for moderate $Re_\lambda \gtrsim 50$, and for the $Re_\lambda \approx 2,000 - 5,000$, L/SES would require $\sim 10^6$ times fewer cells than a DNS.

At the same time, the cell ratio in eq. (29) overestimates the total cost of an L/SES calculation. The SES and DNS have the same resolution, and thus require the same number of time steps to evolve the flow over the same time interval. The LES, however, is much coarser and therefore needs significantly fewer steps in proportion to the ratio of the SES and LES cell sizes, which in turn scales as $\Delta x_S / \Delta x_L \sim \eta/\lambda \sim Re_\lambda^{-1/2}$. Therefore, the ratio of the total number of cell steps, i.e., the number of cells in a grid times the number of computational steps needed to advance a calculation over a given time, in L/SES relative to the DNS scales as

$$\text{Step ratio} \equiv \frac{N_L^3 \frac{\Delta x_S}{\Delta x_L} + N_S^3}{N_D^3} \sim a_\lambda^{-4} Re_\lambda^{-2} + \left(\frac{a_\lambda}{a_l}\right)^3 Re_\lambda^{-3}. \quad (30)$$

For the same reason as with eq. (29), the second term dominates, again leading to the scaling $\sim Re_\lambda^{-3}$ of the cell step ratio, unless $Re_\lambda$ is extremely large, at which point the cell step ratio drops as $\sim Re_\lambda^{-2}$. This scaling of the cell step ratio is demonstrated for the five $Re_\lambda$ cases in Fig. 12 (red squares).

The values of the cell and cell step ratios, along with the grid sizes, for all L/SES calculations discussed here are also listed in Table VI. In particular, in the highest $Re_\lambda \approx 600$ case, the L/SES required 0.24% of cells of the comparable DNS, and only 0.07% of the cell steps, which amounts to a decrease in the computational cost of more than three orders of magnitude.

## VI. CONCLUSIONS

We described *a posteriori* analysis of the Large/Small Eddy Simulation approach (LSES) introduced in Moitro, Dammati, and Poludnenko[9]. The L/SES couples a low-fidelity large-scale LES with a high-fidelity SES aimed at fully resolving the small-scale dynamics in a targeted sub-region of interest. Coupling is achieved by injecting into the SES the filtered and interpolated large-scale velocity field from the LES. In the current formulation, this coupling is one-way. The overall method does not make any assumptions about the flow isotropy or homogeneity, and the only two requirements are (i) that the large scales are accurately captured by the LES, and (ii) the energy transfer in the cascade is predominantly downscale.

As in Moitro, Dammati, and Poludnenko[9], our analysis considers weakly compressible, homogeneous isotropic turbulence (HIT). Here we extended that earlier work, which focused on the *a priori* comparison with the DNS, in three main directions. First, unlike Moitro, Dammati, and Poludnenko[9], which used a DNS flow field to force the SES, all L/SES calculations presented here are based on LES.

Second, we systematically considered all key L/SES parameters, namely the filter scale, $\Delta$, LES grid cell size, $\Delta x_L$, SES domain size, $L_S$, and the time interval of the LES forcing snapshots, $\Delta t_L$, in terms of their effect on the L/SES solution accuracy. Such accuracy was assessed through both global quantities, including the normalized dissipation (dissipative anomaly), and also through various velocity-gradient-based quantities, which reflect complex nonlinear dynamics of the turbulent flow, in particular skewness and higher-order moments of the velocity-gradient statistics. Solution convergence for these metrics was demonstrated for all key L/SES parameters, and specific criteria were formulated for their selection (see § III D).

Third, we analyzed the L/SES solution properties for a range of progressively increasing Taylor-scale Reynolds numbers $Re_\lambda \approx 100 - 600$. Due to the considerable cost of the DNS at the higher $Re_\lambda$, direct comparison with the DNS was performed only at the lowest $Re_\lambda$. For the full range of $Re_\lambda$, the flow field solutions obtained were compared with the published results in the literature.

The normalized dissipation of the turbulent kinetic energy in the L/SES is in agreement with the classical theory of the energy cascade, as the dissipation approaches an asymptotic state at moderate Reynolds numbers $Re_\lambda \gtrsim 200$. The ratios of the integral length scale and the Taylor micro-scale to the Kolmogorov length scale follow the theoretical $Re_\lambda^{3/2}$ and $Re_\lambda^{1/2}$ scalings, respectively. Similarly, the ratio of the root-mean-square velocity fluctuations to the Kolmogorov velocity is also consistent with the theoretical $Re_\lambda^{1/2}$ scaling. Turbulent kinetic energy spectra obtained from the LES and SES pairs of calculations recover both the inertial and dissipation ranges of the turbulent cascade.

We also considered key characteristics, which reflect the small-scale turbulence dynamics. In particular, the skewness of the longitudinal velocity derivatives is close to a constant of $-0.5$ over a wide range of $Re_\lambda$, which is consistent with the previously published results in the literature. In addition, higher-order (even) moments of the velocity derivatives are also investigated. We find close agreement between the L/SES results and the $Re_\lambda$ scalings proposed in the literature for both longitudinal and transverse derivatives.

In addition to the moments of the velocity-gradient statistics, we also considered the full probability density functions (PDF) of enstrophy, $\Omega$, and dissipation, $\epsilon$, from L/SES. In agreement with prior observations in the literature, enstrophy is more intermittent than dissipation with wider PDF and longer tails at all investigated Reynolds numbers. Data from L/SES also show that PDF of both quantities depend on $Re_\lambda$ exhibiting increasing intermittency of the velocity-gradient statistics for larger $Re_\lambda$. Individual PDF were also compared with previously published stretched exponential fits and close agreement was found, especially at higher $Re_\lambda$.

We find that agreement between the L/SES and prior results improves with increasing $Re_\lambda$, with some discrepancy between L/SES and DNS being present at the lowest $Re_\lambda$, especially for the integral scale $l$ and the PDF of $\Omega$ and $\epsilon$. This is a result of the insufficient separation of scales at lower $Re_\lambda$. In particular, lack of separation between the large energy containing scales and the filter scale in the LES suppresses the resulting integral scale. At the same time, proximity of the filter scale and the dissipation scale in the SES introduces numerical errors associated with the filtering of the LES flow field and its subsequent injection into the SES. This affects the small-scale dynamics and leads to a more dissipative SES solution and suppressed intermittency of the velocity-gradient-based quantities.

These results demonstrate the potential of L/SES to capture not only the global quantities and zeroth-order statistics, but also to recover accurately the highly intermittent small-scale behavior, which makes turbulence a challenging phenomenon to study. In particular, for the highest $Re_\lambda \approx 600$, L/SES were able to recover the tails of the PDF of enstrophy and dissipation extending to $\sim 300 - 500$ times the mean value with the probability of $\lesssim 10^{-9}$.

Such extreme events were captured using only 0.24% of cells, and 0.07% of cell-steps, required in an equivalent DNS. We show that the overall cost of the L/SES relative to the DNS drops rapidly as $\propto Re_\lambda^{-3}$. This is due to the fact that the L/SES cost is driven by the $Re_\lambda$ scaling of the ratios $l/\lambda$ and $\lambda/\eta$ separately, both of which grow much more slowly with $Re_\lambda$ than $l/\eta$ that controls the cost of a DNS. Based on this, it can be expected that achieving turbulent regimes with $Re_\lambda$ of a few thousand, which are currently at the forefront of modern exascale computational capabilities using traditional DNS approaches, would incur an almost six orders of magnitude lower computational cost. This would place the exploration of turbulent flows at practically relevant $Re_\lambda$ comparable to those found in the engineering systems and natural flows within reach of modern high-performance computing resources.

Finally, all the L/SES calculations discussed here do not use any explicit subgrid-scale model, thus effectively making them implicit LES.[11] This was done deliberately to study the accuracy of the L/SES under the most conservative assumptions of the lowest LES solution fidelity. The quality of the solution obtained with L/SES, especially at higher $Re_\lambda$, shows that the overall method is not sensitive to the fidelity of the small-scale structures in the LES, provided that large-scale flow dynamics is not affected. At the same time, it would be important to explore further the benefits of explicit subgrid-scale models (along with explicit filtering) in the LES on the overall method accuracy.

In conclusion, the key findings of this study are:

1. L/SES approach can provide solution accuracy comparable to that of a DNS, both in terms of global turbulent characteristics and the small-scale highly intermittent turbulent dynamics, for $Re_\lambda \gtrsim 200$.
2. High solution accuracy can be achieved for larger $Re_\lambda$ even with one-way coupling between the LES and SES and also in the absence of any explicit subgrid-scale models in the LES, which simplifies the implementation of the method and makes it more versatile.
3. L/SES becomes more computationally efficient than DNS for $Re_\lambda \gtrsim 50 - 100$ (depending on the performance of the filtering equation solver). For $Re_\lambda \gtrsim 200$, L/SES cost drops below 1% of a comparable DNS and it decreases as $Re_\lambda{}^{-3}$ beyond that.

## DATA AVAILABILITY

Raw data were generated at the Department of Defense High Performance Computing Modernization Program facilities. Derived data supporting the findings of this study are available from the corresponding author upon reasonable request.

## AUTHOR DECLARATIONS

### Author contributions

**Chang Hsin Chen**: Conceptualization (equal); Data curation (lead); Formal analysis (lead); Investigation (equal); Methodology (equal); Software (supporting); Writing – original draft (equal). **Arnab Moitro**: Conceptualization (equal); Data curation (supporting); Formal analysis (supporting); Investigation (equal); Methodology (equal); Software (lead); Validation (lead); Writing – review & editing (equal). **Alexei Poludnenko**: Conceptualization (equal); Funding acquisition; Supervision; Methodology (equal); Writing – original draft (equal); Writing – review & editing (equal).

### Conflict of interest

The authors have no conflicts to disclose.

## ACKNOWLEDGMENTS

Authors acknowledge funding support by the Air Force Office of Scientific Research under award FA9550-21-1-0012, NASA under award 80NSSC22K0630, and the Office of Naval Research under MURI award N00014-22-1-2606. Computing resources were provided by the Department of Defense High Performance Computing Modernization Program and the US Naval Research Laboratory.